\documentstyle[11pt,epsfig]{article}

\def\laq{~\raise 0.4ex\hbox{$<$}\kern -0.8em\lower 0.62
ex\hbox{$\sim$}~}
\def\gaq{~\raise 0.4ex\hbox{$>$}\kern -0.7em\lower 0.62
ex\hbox{$\sim$}~}

\def\beq{\begin{equation}}
\def\eeq{\end{equation}}
\def\bea{\begin{eqnarray}}
\def\eea{\end{eqnarray}}
\def\bean{\begin{eqnarray*}}
\def\eean{\end{eqnarray*}}

\def\l {\langle}
\def\re {\rangle}

\def \pa {\partial}

\def \ti {\widetilde}

\def \da {\delta}
\def \b {\beta}
\def \a {\alpha}

\def \ga {\gamma}
\def \sg {\sigma}
\def \da {\delta}
\def \ep {\epsilon}
\def \r {\rho}

\def \Om {\Omega}
\def \noi {\noindent}

\def \cR {\cal{R}}

    \def\be{\begin{equation}}
    \def\ee{\end{equation}}
    \def\ba{\begin{eqnarray}}
    \def\ea{\end{eqnarray}}

\begin{document}

\begin{titlepage}

\begin{flushright}
BA-TH/623-09\\
CERN-PH-TH/2009-249\\
arXiv:0912.3244
\end{flushright}

\vspace{0.5cm}

\begin{center}

\huge
{A covariant and gauge invariant formulation \\of  the cosmological  ``backreaction''}

\vspace{0.8cm}

\large{M. Gasperini$^{1,2}$, G. Marozzi$^{3}$ and G. Veneziano$^{4,5}$}

\normalsize

\vspace{0.5cm}

{\sl $^1$Dipartimento di Fisica, Universit\`a di Bari, \\ 
Via G. Amendola 173, 70126 Bari, Italy}

\vspace{.1in}

{\sl $^2$Istituto Nazionale di Fisica Nucleare, Sezione di Bari, \\
Via G. Amendola 173, 70126 Bari, Italy}

\vspace{.1in}

{\sl 
$^3$ GR$\varepsilon$CO --  
Institut d'Astrophysique de Paris, UMR7095,
 CNRS,  \\ 
 Universit\'e Pierre \& Marie Curie, 98 bis boulevard Arago, 75014 Paris, France
}

\vspace{.1in}

{\sl $^4$CERN, Theory Unit, Physics Department, \\ CH-1211 Geneva 23, Switzerland}

\vspace{.1in}

{\sl $^5$Coll\`ege de France, 11 Place M. Berthelot, 75005 Paris, France}

\vspace*{1cm}

\begin{abstract}
\noi
Using  our recent proposal for defining gauge invariant averages we give a general-covariant formulation of  the so-called cosmological
 ``backreaction''. Our effective covariant equations allow to describe  
in an explicitly gauge invariant form the way classical or quantum inhomogeneities affect the average evolution of our Universe.
\end{abstract}
\end{center}

\end{titlepage}

\newpage

\parskip 0.2cm

\section{Introduction}
\label{Sec1}
\setcounter{equation}{0}

It is well known that the homogeneous and isotropic 
Friedman-Lemaitre-Robertson-Walker (FLRW) metric describes the geometric properties of our Universe only on sufficiently large scales of distance, and has to be interpreted as the ``averaged'' cosmological metric: namely, the metric 
emerging from an appropriate smoothing-out of the local inhomogeneities and anisotropies. The same interpretation of averaged variables has to be assigned to the matter energy and momentum density, sourcing the FLRW metric in the cosmological Einstein equations. 

A problem (a rather old one, see for instance \cite{1}) thus appears due to the fact that the Einstein equations for the {\em averaged geometry} are different, in general, from the {\em averaged Einstein equations}. This is because the averaging procedure does not commute, in general, with the non-linear differential operators appearing in  Einstein's equations. As a consequence, the dynamics of the averaged geometry is affected by so-called ``backreaction'' terms, originating from the  contribution of the inhomogeneities present in the metric and matter sectors. 

Interest in these themes has considerably risen after the suggestion that the cosmic acceleration, recently observed on large scales, could be unrelated to phantomatic dark-energy sources, but -- perhaps more simply --  to the dynamical effects of the backreaction (see e.g. \cite{2}) thus solving the well-known``coincidence problem''. This possibility has focused current researches on a new problem: the gauge invariance of the averaging procedure. Indeed, in the absence of gauge invariance, the computed backreaction effects depend not only on the  hypersurface chosen to compute the average integrals, but also on the chosen coordinate frame (see e.g. \cite{LS}-\cite{6} for recent discussions). 

At present, the most commonly used averaging procedure in a cosmological context
is based on the foliation of space-time into three-dimensional
hypersurfaces comoving with the matter sources, and on the volume
integration over such spacelike hypersurfaces \cite{Buchert} (see
\cite{B1} for a recent review, and \cite{6,Brown} for an extension to more general hypersurfaces). 
This procedure is applied, in particular, to the scalar part of the
cosmological Einstein equations, i.e. to  the so-called Hamiltonian
constraint and Raychaudhuri's equation. By spatially averaging such
equations on a comoving domain 
$D$  (and considering, for simplicity, dust fluid sources) one obtains  \cite{Buchert}:
\beq
\left( \dot a_D\over a_D\right)^2= {8 \pi G\over 3} \l \r \re_D -{1\over 6} \left(  \l Q \re_D +\l \cR \re_D\right),
\label{11}
\eeq
\beq
-{ \ddot a_D\over a_D}= {4 \pi G\over 3} \l \r \re_D -{1\over 3}  \l Q \re_D .
\label{12}
\eeq
Here the dot denotes derivatives with respect to the cosmic time of the 
comoving synchronous coordinates, $a_D$ is an effective scale factor,
related to the volume $V_D$ of the integration domain (normalized by a reference volume scale $V_{D_0}$) by 
$a_D = (V_D/V_{D_0})^{1/3}$, where
\beq
V_D= \int_D d^3 x \sqrt{|\det g_{ij}|},
\label{13}
\eeq
and $g_{ij}$ is the intrinsic metric of the comoving hypersurfaces. Also, the brackets denote spatial average over $D$, namely
\beq 
\l \cdots \re_D = {1\over V_D} \int_D d^3 x \sqrt{|\det g_{ij}|} \left( \cdots\right),
\label{14}
\eeq
$\rho$ is the energy density of the dust sources, and $\cR$ is the
scalar intrinsic curvature associated with the spatial metric $g_{ij}$.   
Finally, $\langle Q \rangle$ is a correction called ``kinematical''
backreaction, arising in a particular gauge (see section  \ref{Sec2}) from the
averages of two scalar quantities: the
trace $\Theta$ of the 
expansion tensor and the scalar shear $\sg^2$:
\beq
 \l Q \re_D={2\over 3} \left(  \l \Theta^2 \re_D-  \l \Theta \re_D^2 \right)-
 2  \l \sg^2 \re_D.
 \label{15}
 \eeq
 
The above averaged equations are obtained within a spatial slicing of
the space-time manifold induced by the flow lines of the matter
sources. In this paper we present a covariant 
version of the effective equations for the averaged cosmological quantities
(called hereafter, for simplicity, averaged cosmological equations),
based on
a recently proposed gauge invariant
averaging prescription \cite{GMV1}: the backreaction effects we obtain
depend on the  hypersurface chosen to define the physical observer, 
but {\em do not depend} on the particular
choice of coordinates. In the appropriate class of gauges we recover the recent results given in \cite{6,Brown}. In addition, for an  observer at rest with respect to the matter sources, we recover the same results
as in \cite{Buchert, B1}. 

The paper is organized as follows. In  Sect. \ref{Sec2} we present an explicitly covariant version of the Hamiltonian constraint, Raychaudhuri's equation, and of the projected conservation equation  for a generic foliation of spacetime and energy-momentum tensor. In Sect. \ref{Sec3} we briefly summarize the main aspects of our gauge-invariant averaging prescription, and  we derive the corresponding general-covariant version of the averaged cosmological equations. Our conclusive remarks are briefly presented in Sect. \ref{Sec4}.


\section {Covariant ADM equations}
\label{Sec2}
\setcounter{equation}{0}
In order to introduce spatial averages of physical quantities we  consider
a general class of foliations of spacetime by spacelike
hypersurfaces $\Sigma(A)$ over which a scalar field $A(x)$ takes
constant values  (the so-called level-sets of $A$). Let $n_{\mu}$ be the
future-directed unit normal to 
$\Sigma(A)$, defined by 
\begin{equation}
\label{2.1}
n_{\mu} = -  \frac{\partial_{\mu} A}{ (-
 \partial_{\mu} A \partial_{\nu} A ~ g^{\mu\nu}) ^{1/2}} \; \;, 
 \,~~~~~~~~~~~~~~ \, n_{\mu} n^{\mu} = -1
\label{DefnA}
\end{equation} 
(we are using the metric signature $(-,+,+,+)$).
Let us also introduce the projector $h_{\mu\nu}$ into the hypersurfaces by:
\begin{equation}
h_{\mu\nu} = g_{\mu\nu} + n_{\mu}n_{\nu}\,,~~~~~~~~~
h_{\mu\rho}h^{\rho}_{\nu} = h_{\mu\nu}\, , ~~~~~~~~~
h_{\mu\nu}n^{\mu} = 0\,.
\end{equation} 
The Einstein equations $G_{\mu\nu} = T_{\mu\nu}$ (we use units in which $8 \pi G =1$) can then be projected along $n^{\mu}$ and $h^{\mu\nu}$, and give rise to three (sets of) equations that can be chosen in the following form:
\bea
G_{\mu\nu} n^{\mu}n^{\nu} &= &T_{\mu\nu} n^{\mu}n^{\nu} \equiv \varepsilon \; , \label{2.3} \\
G_{\mu\nu} n^{\mu}h^{\nu}_{\rho} &=& T_{\mu\nu} n^{\mu}h^{\nu}_{\rho}  \equiv J_{\rho} \; , \label{2.4} \\
R_{\mu\nu} h^{\mu}_{\rho}h^{\nu}_{\sigma} &=& T_{\mu\nu} h^{\mu}_{\rho}h^{\nu}_{\sigma} -\frac12 h_{\rho\sigma} T \equiv  {\cal S}_{\rho\sigma} 
 -\frac12 h_{\rho\sigma} T \, . \label{2.5}
\eea
They correspond to an explicitly covariant version of the so-called  Arnowitt-Deser-Misner (ADM) equations. 

It is always possible to make contact with the more conventional ADM formalism by choosing a class of gauges   in
which the scalar field $A(x)$  is homogeneous. We shall call it the ADM gauges.
In such a gauge the normal  vector $n_{\mu}$ takes the  form:
\begin{equation}
\label{2.6}
n_{\mu} = N (-1,0,0,0)\;, ~~~~~~~~~~ n^{\mu} = {1\over N}(1,-N^i )\;,
\end{equation} 
where $N$ and $N^i$ are, respectively, the so-called lapse function and shift vector. In this gauge:
\begin{equation}
h^{\mu}_i = \delta^{\mu}_i\; , ~~~~~~~~~~~~ h^{0}_0 = 0  \;, \; ~~~~~~~~~~~~ h_{ij} = g_{ij}\;,
\end{equation} 
where  $ g_{ij}$ is the induced 3--metric  (or first fundamental form) on $\Sigma$.
The spacetime metric  of the foliated spacetime is  given by:
\begin{equation}
ds^2 = g_{\mu\nu} dx^{\mu} dx^{\nu} = 
-N^2 dt^2 + g_{ij} (dx^i + N^i dt)(dx^j + N^j dt) \;\;\;,
\end{equation}
and its inverse by:
\begin{equation}
\Box \equiv g^{\mu\nu} \nabla_\mu \nabla_\nu = 
-N^{-2}(\nabla_0 - N^i \nabla_i)^2  + {}^{(3)}\!g^{ij} \nabla_i \nabla_j \;\;\;.
\end{equation}
where $^{(3)}\!g^{ij} $ is the inverse of the $3 \times 3$ metric $g_{ij}$.

In this class of ADM gauges Eqs. (\ref{2.3}) and (\ref{2.4}) reduce to the convential form of the Hamiltonian and momentum constraints, respectively, while Eq. (\ref{2.5}) generates the second order evolution equations. In the ADM context such evolution equations are splitted into twice as many  equations,  those defining the second fundamental form (or extrinsic curvature)  $K_i^j$, and those describing the evolution of $K_i^j$ itself (see, e.g. \cite{B1}).

In order to give a covariant and gauge invariant formulation of the cosmological
backreaction we shall make use of the covariant Eqs. (\ref{2.3})--(\ref{2.5}),  which lead to an explicitly scalar form of the Hamiltonian constraint and  Rachayduri's equation.
To this purpose, let us first consider the spacetime flow generated by the timelike vector field $n_\mu$, and define the projected expansion tensor of the flow worldlines as
\be
\Theta_{\mu\nu}\equiv h^\alpha_\mu h^\beta_\nu \nabla_\a n_\b
=\frac{1}{3} h_{\mu\nu}\Theta+\sigma_{\mu\nu}+\omega_{\mu\nu}\; ,
\ee
where
\be 
\Theta \equiv \nabla_\mu n^\mu, ~~~~~ \sigma_{\mu\nu}\equiv h^\alpha_\mu
h^\beta_\nu \left(\nabla_{(\a} n_{\b)}-\frac{1}{3}h_{\alpha\beta}
\nabla_\tau n^\tau \right), ~~~ \omega_{\mu\nu}\equiv
h^\alpha_\mu h^\beta_\nu \nabla_{[\a} n_{\b]},
\ee
 are the expansion scalar, the
shear tensor and the rotation tensor, respectively.
In our case, with $n_{\mu}$ given by Eq. (\ref{2.1}), we have
a zero rotation tensor and we can write 
\be
\sigma_{\mu\nu}=\Theta_{\mu\nu}-\frac{1}{3} h_{\mu\nu}\Theta \, , ~~~~~~~ \sigma^2\equiv\frac{1}{2} \sigma^\mu_\nu \sigma^\nu_\mu=
\frac{1}{2} \left(\Theta^\mu_\nu \Theta^\nu_\mu-\frac{1}{3}
\Theta^2\right)\, ,
\label{sigma2}
\ee
where $\sigma^2$ is the shear scalar.

Consider first the Hamiltonian constraint (\ref{2.3}), whose left hand side can be decomposed as follows:
\beq
R_{\mu\nu}n^\mu n^\nu+\frac12 R = \frac12 \left( \Theta^2 - \Theta^\mu_\nu 
\Theta^\nu_\mu\right) + \frac12 {\cal R}_s\; .
\eeq
 In the ADM gauge the term in brackets on the right hand side can be expressed in terms of the extrinsic curvature as:
\beq
\Theta^2-\Theta^\mu_\nu 
\Theta^\nu_\mu=K^2-K^i_j K^j_i\, ,
\eeq
while ${\cal R}_s$ is a  scalar which, in the ADM gauge, goes over to the intrinsic scalar curvature ${\cal R}$ associated with the induced metric $g_{ij}$.
We can thus rewrite the Hamiltonian constraint completely in terms of scalar quantities as:
\beq
\label{H}
{\cal R}_s+\Theta^2-\Theta^\mu_\nu\Theta^\nu_\mu= {\cal R}_s+\frac{2}{3}\Theta^2-2\sigma^2 = 2 T_{\mu\nu} n^\mu n^\nu = 2  \varepsilon\, .
\label{215}
\eeq

Coming now to the explicitly scalar form of Rachayduri's equation, we note that it corresponds to the linear combination of Eq. (\ref{2.3}) and of the trace of Eq. (\ref{2.5}), leading to:
\be 
R_{\mu\nu}n^\mu n^\nu = T_{\mu\nu} h^{\mu\nu}-\frac12 T. 
\ee
After some straightforward calculation the above equation takes the form:
\bea
-n^\mu \nabla_\mu \Theta&=& 2 \sigma^2 + \frac{1}{3} \Theta^2 - \nabla^\nu(n^\mu\nabla_\mu n_\nu) + \left(
T_{\mu\nu}-\frac{1}{2}g_{\mu\nu} T\right) n^\mu n^\nu  \nonumber \\ 
&=& 2 \sigma^2 + \frac{1}{3} \Theta^2 - \nabla^\nu(n^\mu\nabla_\mu n_\nu) + \varepsilon
 + \frac12  T\,.
\label{Raychaudhuri}
\eea
 
 So far we have made no assumptions on the form of $T_{\mu\nu}$.
Let us now restrict ourselves to the case of a perfect fluid with:
\be
 T_{\mu\nu}=(\rho+p)u_\mu u_\nu +p g_{\mu\nu},
 \ee
  where $u_\mu$ is the
4-velocity comoving with the fluid, and $\rho$ and $p$ are, respectively, the (scalar) energy density and pressure in the  fluid's rest frame (the generalization to several non-interacting fluids is straightforward). We stress that $u_{\mu}$ and $n_{\mu}$ are in general distinct: the former depends on the properties of the matter sources, the latter depends on the choice of the hypersurfaces on which we want to average, hence should be determined by the particular problem at hand. With this model of sources we can then express the basic quantities entering our equations (\ref{H}) and (\ref{Raychaudhuri}) as:
\be
\varepsilon=T_{\mu\nu} n^\mu n^\nu=(\rho+p)(u^\mu n_\mu)^2-p \; , ~~~~~~~~~~ T =T_{\mu}^{\mu} = - \rho+3 p \; .
\ee

For later use, we note that the trace $T$ of the energy momentum tensor  is unaffected by the possible ``tilt'' (misalignment) between $u_{\mu}$ and $n_{\mu}$, while 
the sources of the covariant ADM Eqs. (\ref{2.3})--(\ref{2.5}) -- namely the objects that we may call the ADM energy density $\varepsilon$, the ADM pressure $\pi = {\cal S}_{\rho}^{\rho}/3$, and the ADM current $J_{\nu}$ -- become:
\bea
\label{2.19}
\varepsilon &=& \rho - (\rho+p)\left(1- (u^\mu n_\mu)^2 \right)\; , \
\\ 
\pi &=& p - \frac13 (\rho+p) \left(1- (u^\mu n_\mu)^2\right)\;,
\label{2.20} \\
J_{\nu} &=&  (\rho+p)  u^\b n_\b(1+ u^\mu n_\mu) u_{\nu}.
\label{2.21} 
\eea
On the other hand, a straightforward calculation in the ADM gauge leads to:
\be
\label{2.22}
(u^\mu n_\mu)^2 = 1 +  {}^{(3)}\!g^{ij}u_i u_j \ge 1\,,
\ee
meaning that this quantity is always larger than 1. We can thus introduce a ``tilt angle'' $\alpha_{T}$
such that $\sinh^2 \alpha_{T}=(u^\mu n_\mu)^2 -1$, and we can rewrite $\ep$ and $\pi$ in the more convenient form
\be
\varepsilon = \rho + (\rho+p) \sinh^2 \alpha_{T}\, , ~~~~~~~~~{\bf \pi} = p + \frac{1}{3}(\rho+p) \,
\sinh^2 \alpha_{T}\,.
 \label{KeyDE}
\ee

Let us finally consider the condition following from the projected energy-momentum conservation law, $n^\mu \nabla^\mu T_{\mu\nu}=0$. It is known that such condition is not implied by the two projected scalar Einstein equations considered above, and has to be added to the set of averaged cosmological equations as an additional independent constraint \cite{Buchert,B1}. In the general context we are considering (with $n_\mu \not= u_\mu$), the projected conservation equation can be written explicitly as follows:
\beq
u^\mu \pa_\mu \left[\left( \r + p \right) u^\r n_\r \right]+ n^\mu \pa_\mu p
+ \left( \r + p \right) \left[ \nabla_\mu u^\mu u^\r n_\r - \Theta^{\mu\nu} u_\mu u_\nu \right]=0. 
\label{2.25}
\eeq

\section{Covariant averaged equations}
\label{Sec3}
\setcounter{equation}{0}

Let us begin from the four dimensional integral of a scalar $S(x)$ as
defined in \cite{GMV1}:
\beq
I(S,\Om) = \int_{\Om(x)} d^{4} x  \sqrt{-g(x)} \,S(x) \equiv
 \int_{{\cal M}_4} d^{4} x  \sqrt{-g(x)} \,S(x)W_\Om(x).
\label{3.1}
\eeq
The integration region $\Om \subseteq {\cal M}_4 $ is defined in terms of a suitable scalar window function
$W_\Om$, selecting a region with temporal boundaries determined by the space-like hypersurfaces  $\Sigma (A)$ (defined in Sect. \ref{Sec2}), and with spatial boundary determined by the coordinate condition $B<r_0$, where $B$ is a (positive) function of the coordinates with space-like gradient $\pa_\mu B$, and $r_0$ is a positive constant. As we are interested in the variation of the volume averages along the flow lines normal to $\Sigma (A)$, we choose in particular the following window function:
\beq
W_\Omega(x)=
n^{\mu} \nabla_\mu \theta(A(x)-A_0) 
\theta(r_0-B(x)),
\label{3.2}
\eeq
where $\theta$ is the Heaviside step function, and $n_{\mu}$ is defined in Eq. (\ref{2.1}).

As discussed in \cite{GMV1}, if $B(x)$ is a scalar function the integral (\ref{3.1}) is not only a scalar under general coordinate transformations but is also gauge invariant (to all orders): namely, it is invariant under the local field reparametrizations induced by any coordinate change when old and new fields are evaluated at the same space-time position. If $B$ is not a scalar, the spatial boundary can be a source of
breaking of  covariance and gauge invariance. In \cite{GMV2} we will discuss in more
detail such a breaking, and confirm that it goes away in the limit of
large spatial volumes (with respect to the typical scale of 
inhomogeneities) \cite{GMV1}. Using the window function (\ref{3.2}),  the integral (\ref{3.1}) becomes:
\beq
I(S,A_0) = 
 \int_{{\cal M}_4} d^{4} x  \sqrt{-g(x)}  \delta(A(x)-A_0) (-\partial_\mu A \, \partial^\mu A)^{1/2}\theta(r_0-B(x)) \,S(x)\,.
\label{3.3}
\eeq

Let us now consider  the derivative of $I(S,A_0)$ with respect to  $A_0$, a
quantity that, like $I$ itself, is covariant and gauge invariant (apart from
a possible gauge dependence induced by the spatial boundary): 
\begin{eqnarray}
\frac{\pa I(S,A_0)}{\pa A_0}= && 
\!\!\!\!\!\!\!\!\!\!
-  \int d^{4} x  \sqrt{-g(x)} \, \delta'\left(A(x)-A_0\right) (-\partial_\mu A \, \partial^\mu A )^{1/2}\theta(r_0-B(x)) \,S(x)  \nonumber \\
=&& \!\!\!\!\!\!\!\!\!\!
- \int d^{4} x  \sqrt{-g(x)}\, \pa_0 \delta \left(A(x)-A_0\right) 
\left[\pa_0A(x)\right]^{-1} (-\partial_\mu A \, 
\partial^\mu A)^{1/2}\theta(r_0-B(x)) \,S(x).  
\nonumber \\ &&
\end{eqnarray}
Within adapted ADM coordinates, where $A$ is homogeneous, we can always 
choose those with vanishing shift (i.e. with $g_{00}=-N^2$ and $g_{0i}=0$).  In such coordinates we have:
\begin{eqnarray}
\frac{\pa I(S,A_0)}{\pa A_0}
=&& \!\!\!\!\!\!\!\!\!\! - \int d^{4} {x}  \sqrt{-{g}}~\pa_0
\delta \left(A({t})
-A_0 \right) (-{g}^{00})^{1/2} \theta(r_0-B(x))\,{S}({x}) 
  \cr
=&& \!\!\!\!\!\!\!\!\!\! \int d^{4} {x}  \,
\delta \left(A({t})
-A_0 \right) \pa_0\left[ \sqrt{-{g}} (-{g}^{00})^{1/2} \theta(r_0-B(x))\,{S}({x}) 
 \right] \cr
=&&  \!\!\!\!\!\!\!\!\!\! \int d^{4}{x}\, \sqrt{|{\gamma}|}  
\delta(A({t})-A_0)
\Big[\theta(r_0-B(x))\left(N{\Theta}\,{S}+\pa_0{{S}}\right) -
\da (r_0-B(x)) S \pa_0B \Big],
\cr &&
\label{Ider}
\end{eqnarray}
where $\ga= \det g_{ij}$, and we have used that, in these coordinates,
\be
\Theta   = N^{-1} \partial_0 \log \sqrt{\gamma} .
\label{Theta}
\ee
The above equation 
can be easily recast into the following covariant and gauge invariant form, 
 \be
\frac{\pa I(S,A_0)}{\pa A_0} = I\left(\frac{\partial_{\mu} A \partial^{\mu} S}{
\partial_{\mu} A \partial^{\mu} A},  A_0\right) + I \left(  \frac{S \, \Theta}{(-\partial_{\mu} A \partial^{\mu}
A)^{1/2}} , A_0 \right)-2 I\left(\frac{\partial_{\mu} A \partial^{\mu} B}{
\partial_{\mu} A \partial^{\mu} A}  S \da(r_0-B),  A_0\right),
\label{Idercov} 
\ee
that reduces to Eq. (\ref{Ider})  in the special  coordinates we have been using. Note that the last factor of the above equation is absent if $n^\mu$ is orthogonal to the gradient of $B$   ($n^\mu \pa_\mu B=0$), namely if $B$ does not depend on the time coordinate of the gauge used in Eq. (\ref{Ider}). We shall restrict ourselves to this case hereafter.
 
Let us define now the covariant averaging prescription for a scalar $S(x)$ on the hypersurfaces of constant $A$, following \cite{GMV1}, as:
\beq
\langle S \rangle_{A_0} = \frac{I(S, A_0)}{I(1, A_0)}. 
\label{defav}
\eeq
Using Eq. (\ref{Ider}) we can easily obtain the derivative of the averaged scalar $\l S \re_{A_0}$ as:
\beq
\frac{\pa \langle S \rangle_{A_0}}{\pa A_0} = \langle \partial_{A_0} S \rangle_{A_0} + \left\langle S\frac{N \Theta}{\pa_0{A}} \right \rangle_{A_0} - 
 \langle S  \rangle_{A_0}  \left \langle \frac{N \Theta}{\pa_0{A}} \right \rangle_{A_0}\, ,
\label{derav}
\eeq
and  Eq. (\ref{Idercov}) immediately gives us the generally covariant version of the above equation:
\be
\frac{\partial \langle S \rangle_{A_0}}{\partial A_0} = \left\langle 
\frac{\partial_{\mu} A \partial^{\mu} S}{
\partial_{\mu} A \partial^{\mu} A} \right\rangle_{A_0}+ 
\left\langle \frac{S \, \Theta}{(-\partial_{\mu} A \partial^{\mu}
A)^{1/2}} 
\right\rangle_{A_0} - 
 \langle S  \rangle_{A_0}  
\left\langle\frac{\Theta}{(-\partial_{\mu} A \partial^{\mu}
A)^{1/2}} 
\right\rangle_{A_0} .
\label{gideravB}
\ee
This is the covariant and gauge invariant generalization of the Buchert-Ehlers 
commutation rule \cite{BE}, and will be the starting point for our generalization of the  averaged cosmological equations. However, let us first illustrate the precise connection to the special version of this rule obtained in \cite{BE}.

In ADM coordinates, 
with $A$ homogeneous, Eq. (\ref{3.3}) reads:
\be
I(S,A_0) = \int_{\Sigma_{A_0}}  d^3 x  \sqrt{|{\gamma}(t_0, \vec{x})|} 
\, {S}(t_0, \vec{x}) \theta\left(r_0-B(t_0, \vec x)\right)\,,
\ee
where we have called $t_0$ the time  when $A({t})$ takes the
constant values $A_0$, and the averages are referred to a section of the three-dimensional hypersurface $\Sigma_{A_0}$ where $A(x)=A_0$. This is exactly the type of spatial integrals used in \cite{Buchert} (with a domain $D$ determined  by the condition $B(x) <r_0$), and thus leads to the same definition of averages (see Eq. (\ref{14})). Let us compute, in this case, the corresponding version of Eq. (\ref{gideravB}). Multiplying both sides by $\pa_t A_0$ we obtain the equation
\bea
\pa_t \langle S \rangle_{A_0}&=& \left \l  \pa_t S -N^i \pa_i S
\right \re_{A_0} +\left\langle S{N \Theta} \right \rangle_{A_0} - 
 \langle S  \rangle_{A_0}  \left \langle {N \Theta} \right \rangle_{A_0} \cr
 \vspace{.2in}
 &=& \left \l  N n^{\mu} \pa_{\mu} S 
\right \re_{A_0} +\left\langle S{N \Theta} \right \rangle_{A_0} - 
 \langle S  \rangle_{A_0}  \left \langle {N \Theta} \right \rangle_{A_0} ,
\eea
which generalises the commutation rule given in \cite{BE} to the case of non-vanishing shift vector. For $N^i=0$ the standard result is recovered. 

Let us come now to the covariant formulation of the averaged cosmological equations. Starting from the generally covariant volume integral
\be
I (1, A_0)=\int d^4 x \sqrt{-g}\, \delta(A(x)-A_0)
\sqrt{- \partial_\mu A \partial^\mu A}\, \theta(r_0-B(x)) 
\ee
we define, along the lines of the previous approach \cite{Buchert}, an effective scale factor $\ti a$ such that
\be
\frac{1}{\ti a} \frac{\partial \, \ti a}{\partial A_0}= 
 \frac{1}{3I (1, A_0)} \frac{\partial \, I (1, A_0)}{\partial A_0}.
\ee
We then find, using Eqs. (\ref{Idercov}), (\ref{defav}), 
\be 
\frac{1}{\ti a} \frac{\partial \, \ti a}{\partial A_0}=\frac{1}{3} \left\langle\frac{\Theta}{(-\partial_{\mu} A \partial^{\mu}
A)^{1/2}} 
\right\rangle_{A_0}\,.
\label{HD}
\ee
We are now in the position of presenting the covariant generalization of
the averaged equation (\ref{11}). Taking the square of the previous
equation, using the Hamiltonian constraint in the form of
Eq. (\ref{215}), 
and explicitly reintroducing Newton's constant in the formulae, we easily
obtain:
\bea
\left(\frac{1}{{\ti a}}\frac{\partial \, \ti a}{\partial A_0} \right)^2 &=&
\frac{8\pi G}{3} \left\langle\frac{\varepsilon}{(-\partial_{\mu} A \partial^{\mu}
A)} \right\rangle_{A_0}-\frac{1}{6}
\left\langle\frac{{\cal R}_s}{(-\partial_{\mu} 
A \partial^{\mu}
A)} \right\rangle_{A_0} 
\nonumber \\
& & 
\!\!\!\!\!\!\!\!\!\!\!\!\!\!\!\!\!\!\!\!\!\!\!\!\!\!\!\!\!\!\!\!\!\!\!\!\!
-\frac{1}{9}\left[   
\left\langle\frac{\Theta^2}{(-\partial_{\mu} A \partial^{\mu}
A)} \right\rangle_{A_0}
-\left\langle\frac{\Theta}{(-\partial_{\mu} A \partial^{\mu}
A)^{1/2}} 
\right\rangle_{A_0}^2 \right]
+\frac{1}{3} \left\langle\frac{\sigma^2}{(-\partial_{\mu} A \partial^{\mu}
A)} \right\rangle_{A_0}.
\label{EQB2}
\eea

In order to arrive at the covariant generalization of the second Buchert's equation (\ref{12}) we start with the simple relation
\beq
{1\over \ti a}{\pa^2 \ti a \over \pa A_0^2}= 
{\pa \over \pa A_0}\left({1\over \ti a}{\pa \ti a \over \pa A_0}\right)+
\left({1\over \ti a}{\pa \ti a \over \pa A_0}\right)^2.
\label{317}
\eeq
Using Eq. (\ref{HD}), and the general commutation rule (\ref{gideravB}) for the first term on the right hand side, we obtain
\bea
-\frac{1}{\ti a} \frac{\partial^2 \, \ti a}{\partial A_0^2}&=&
\frac{2}{9}  \left\langle\frac{\Theta}{(-\partial_{\mu} A \partial^{\mu}
A)^{1/2}} 
\right\rangle_{A_0}^2+\frac{1}{6} \left\langle \frac{\partial_\mu A 
\partial^\mu (
\partial_\nu A\partial^\nu A)}{(-\partial_\mu
A\partial^\mu A)^{5/2}} \Theta \right\rangle_{A_0}
\nonumber \\
& & 
-\frac{1}{3} \left\langle\frac{\Theta^2}{(-\partial_{\mu} A \partial^{\mu}
A)} \right\rangle_{A_0}+\frac{1}{3} \left\langle\frac{\partial_\mu A 
\partial^\mu \Theta}{(-\partial_\mu
A\partial^\mu A)^{3/2}}\right\rangle_{A_0}.
\label{D2aD}
\eea
Inserting then the covariant Raychaudhuri's equation (\ref{Raychaudhuri}) in the last term of the above equation we are lead to the final result,
\bea
-\frac{1}{\ti a} \frac{\partial^2 \, \ti a}{\partial A_0^2}&=&
\frac{4 \pi G}{3} \left\langle\frac{\varepsilon+3 {\pi}}{(-\partial_{\mu} A \partial^{\mu}
A)} \right\rangle_{A_0}-\frac{1}{3}\left\langle
\frac{\nabla^\nu(n^\mu\nabla_\mu n_\nu)}{(-\partial_{\mu} A \partial^{\mu}
A)} \right\rangle_{A_0}
+\frac{1}{6} \left\langle\frac{\partial_\mu A 
\partial^\mu (\partial_\nu A\partial^\nu A)}{(-\partial_\mu
A\partial^\mu A)^{5/2}} \Theta \right\rangle_{A_0}
\nonumber \\
& &
\!\!\!\!\!\!\!\!
-\frac{2}{9}\left[   
\left\langle\frac{\Theta^2}{(-\partial_{\mu} A \partial^{\mu}
A)} \right\rangle_{A_0}
-\left\langle\frac{\Theta}{(-\partial_{\mu} A \partial^{\mu}
A)^{1/2}} 
\right\rangle_{A_0}^2 \right]
+\frac{2}{3} \left\langle\frac{\sigma^2}{(-\partial_{\mu} A \partial^{\mu}
A)} \right\rangle_{A_0},
\label{EQB1}
\eea
where we have used the relation $\rho -3 p= \varepsilon-3\pi$ (see Eq. (\ref{KeyDE})).  Equations (\ref{EQB2}) and (\ref{EQB1}), together with the averaged conservation equation discussed below,  are the main results of this paper.  

Let us now observe that Eq. (\ref{EQB2}), written in the ADM gauge, and multiplied by $(\pa_t A_0)^2$, reduces to the equation recently presented in \cite{6,Brown}:
\bea
\left({1\over \ti a} {\pa \ti a\over \pa t} \right)^2 &=& {8 \pi G\over 3} \l N^2 
\rho \re_{A_0}+
 {8\pi G\over 3} \left \l N^2 
 (\rho+p) \sinh^2 \alpha_{T} \right \re_{A_0}
-{1\over 6} \l N^2 {\cal R} \re_{A_0}
\nonumber \\ &-&
{1\over 9} \left(\l N^2 \Theta^2 \re_{A_0}- \l N \Theta \re_{A_0}^2\right)+ {1\over 3} \l N^2 \sg^2\re_{A_0}. 
\eea
We have used Eq. (\ref{KeyDE}) to replace the ADM parameter $\varepsilon$ with the fluid proper energy and pressure. When $\sinh^2 \alpha_{T}=0$ (namely when the averaging hypersurfaces coincide with those orthogonal to the fluid velocity, $n_\mu=u_\mu$), we exactly recover the corresponding Buchert's equation (see the second paper of  \cite{Buchert}). In the synchronous gauge ($N=1$) and for dust sources ($p=0$) we recover instead Buchert's Eq. (\ref{11}), apart from the additional contribution arising from a nonvaninshing tilt angle  $\alpha_{T}$. 

Consider now the second equation (\ref{EQB1}), and let us first note that 
\beq
\frac{1}{\ti a} \frac{\partial^2 \, \ti a}{\partial t^2}= \left(\frac{\partial A_0}{\partial t}\right)^2
\frac{1}{\ti a} \frac{\partial^2 \, \ti a}{\partial A_0^2}+
\frac{1}{\ti a} \frac{\partial^2 \, A_0}{\partial t^2}{\pa \ti a\over \pa A_0}. 
\label{322}
\eeq
We go then to the ADM coordinates, as before, imposing in addition the convenient gauge choice $N^i=0$,  and we apply Eq. (\ref{KeyDE}) to express $\varepsilon$ and $\pi$ in terms of $\rho$ and $p$. By using Eq. (\ref{EQB1}) to eliminate the first term on the right hand side of Eq. (\ref{322}) we obtain
\bea
-\frac{1}{\ti a} \frac{\partial^2 \, \ti a}{\partial t^2}&=&{4 \pi G\over 3}\left \l N^2 (\rho+ 3 p)\right \re_{A_0}+ {8 \pi G\over 3}\left \l N^2 (\rho+  p)\sinh^2 \a_{T} \right \re_{A_0}
\nonumber \\
&-&{2\over 9} \left(\l N^2 \Theta^2 \re_{A_0}- \l N \Theta \re_{A_0}^2\right)+ {2\over 3} \l N^2 \sg^2\re_{A_0} 
\nonumber \\
&-&{1\over 3}\left \l \Theta {\pa N\over \pa t} \right \re_{A_0}
-{1\over 3}\left \l N ~ {}^{(3)}\!g^{ij} \nabla_i \nabla_j N \right \re_{A_0},
\eea
in agreement with \cite{6}. Again, for $\sinh^2 \alpha_{T}=0$, we also recover the corresponding Buchert's equation (see for instance \cite{B1}). We may note that, when $\sinh^2 \alpha_{T}\not=0$, the ``tilt effects'' give a negative contribution (assuming $\rho+  p > 0$)  to the average cosmic acceleration. 

Let us conclude this section by noting that, as anticipated in Sect. \ref{Sec2}, the set of averaged cosmological equations has to be complemented by the general-covariant average of the conservation equation (\ref{2.25}). Such an operation can be performed straightforwardly, according to the general procedure outlined above. We shall present here, for simplicity, the covariant version of the averaged conservation equation in the particular case in which the space-time foliation is referred to the fluid comoving frame. In such a case, setting $n_\mu= u_\mu$ in Eq. (\ref{2.25}), and applying the 
general-covariant commutation rule (\ref{gideravB}), we obtain
\beq
{\pa \over \pa A_0} \l \r \re_{A_0}= - \left \l {\Theta\, p \over (- \pa_\mu A \pa^\mu A)^{1/2}}  \right \re_{A_0} - \l \r \re \left \l {\Theta \over (- \pa_\mu A \pa^\mu A)^{1/2}}  \right \re_{A_0}
\label{323}
\eeq
(note that, in this case, the scalar $A$ corresponds to the velocity potential of the fluid sources). Going, as before, to the ADM coordinates, and multiplying by $\pa_t A_0$, we recover the known result  
\beq
{\pa \over \pa t} \l \r \re_D+ \left \l {N\Theta\, p }  \right \re_{D} - \l \r \re \left \l N{\Theta }  \right \re_{D}=0,
\eeq
already presented in various papers \cite{Buchert,B1}. 

\section{Conclusion}
\label{Sec4}
\setcounter{equation}{0}

The main results of this paper are the covariant and gauge invariant formulation of the Buchert-Ehlers commutation rule, Eq. (\ref{gideravB}), and of the effective equations for the averaged evolution of a perfect fluid-dominated Universe, Eqs. (\ref{EQB2}), (\ref{EQB1}),(\ref{323}). The average is performed over a generic class of hypersurfaces, not necessarily orthogonal to the fluid flow lines. We stress that our results allow to compute averaged quantities in a completely arbitrary coordinate system.

The results obtained in this paper can be directly applied to the case of the quantum cosmological backreaction, by using the correspondence between quantum expectation values and classical averages performed over all three-dimensional space, as illustrated in details in \cite{GMV1}. Hence, in particular, can be applied to study the effect of the quantum fluctuations within the canonical formalism of cosmological perturbation theory.   

At the same time, the classical averaged equations (possibly extended to light-like hypersurfaces) may provide a covariant starting point for determining whether present inhomogeneities can significantly contribute to the observed cosmic acceleration.

\section*{Acknowledgements}

One of us (MG) is very grateful to the Coll\`ege de France for its warm hospitality and support. One of us (GM) wishes to thank Syksy Rasanen for useful discussions. GM was supported by the GIS ``Physique des Deux Infinis''.



\begin{thebibliography}{999}
\newcommand{\bb}{\bibitem}

\bb{1}G. F. R. Ellis, in ``{\em 10th International Conference on 
General Relativity and Gravitation}'', 
ed. by B. Bertotti, De Felice and A. Pascolini 
(Reidel, Dordrecht, 1984), p. 215.

\bb{2}S. Rasanen, JCAP {\bf 02}, 003 (2004).

\bibitem{LS}
N. Li and D. J. Schwarz, Phys. Rev. {\bf D76}, 083011 (2007).

\bibitem{GMV1}
M. Gasperini, G. Marozzi, and G. Veneziano, JCAP {\bf 03}, 011 (2009).

\bibitem{5}J. Larena,  Phys. Rev. {\bf D79}, 084006 (2009).

\bibitem{6}I. A. Brown, J. Behrend and K. A. Malik, JCAP {\bf 11}, 027 (2009).

\bibitem{Buchert}T. Buchert, Gen. Rel. Grav. {\bf 32}, 105 (2000); {\bf 33}, 1381 (2001).

\bb{B1}T. Buchert, Gen. Rel. Grav. {\bf 40}, 467 (2008).


\bb{Brown}I. A. Brown, G. Robbers and J. Behrend, JCAP {\bf 04}, 016 (2009); I. A. Brown, L. Schrempp and K. Ananda, arXiv:0909.1922 [gr-qc]. 


\bibitem{GMV2}
M. Gasperini, G. Marozzi, and G. Veneziano, in preparation.


\bb{BE} T.  Buchert and J. Ehlers, Astron. Astrophys. {\bf 320}, 1 (1997).


\end{thebibliography}
\end{document}